\documentclass[fleqn]{article}%
\usepackage{amsmath}
\usepackage{harvard}%
\usepackage{amsfonts}%
\usepackage{amssymb}%
\usepackage{graphicx}
\providecommand{\U}[1]{\protect\rule{.1in}{.1in}}
\begin{document}

\title{A Note on the Entropy of Entanglement and Entanglement Swapping Bounds}
\author{\textit{Simon J.D. Phoenix}\\Khalifa University, PO Box 127788, Abu Dhabi, UAE}
\maketitle

\begin{abstract}
Using the information content of correlations between multipartite systems,
together with the notion of partitioning, we show that some general results
about the evolution of correlations in quantum systems can be derived with
only elementary methods. In particular, we show that for 2 quantum systems
\textit{A} and \textit{B}, each comprised of a number of sub-systems, in which
a partition of \textit{A} interacts unitarily with a partition of \textit{B},
then the total correlation can only increase (or remain unchanged) and is
given simply by the sum of the initial correlation and the correlation that
develops as a result of the interaction. We then show that in a 4 qubit
entanglement swapping process the transferred degree of entanglement is
bounded by the lower of the initial degrees of entanglements of the qubits.

\end{abstract}

\section{Introduction}

For 2 quantum sub-systems, which we label as $1$ and $2$, the degree of
entanglement between them is well-characterized by the entropy of entanglement
[1]. It is related to the mutual information $I_{12}$, and for pure entangled
states the entropy of entanglement is simply just half this quantity. If
$\rho$ is the total density operator for the $\left\{  12\right\}  $ joint
system and $\rho_{k}$ is the reduced density operator for sub-system $k$, then
the entropy of an individual system is $S_{k}=~$Tr$_{k}\left\{  \rho_{k}%
\ln\rho_{k}\right\}  $ with $\rho_{k}=~$Tr$_{j\neq k}\left\{  \rho\right\}  $
and the total entropy is $S_{12}=~$Tr$\left\{  \rho\ln\rho\right\}  $ then the
mutual information is given by [2]%

\begin{equation}
I_{12}=S_{1}+S_{2}-S_{12}%
\end{equation}
In previous work we termed this quantity the `index of correlation'.
Physically it is a basis-independent measure of the information contained in
the correlations between 1 and 2.

If we wish a measure of correlation for quantum systems to be
basis-independent and additive then the mutual information, or index of
correlation, is the unique measure satisfying these properties [3]. By
additive here it is meant that if 2 systems $A$ and $B$, which could each be
comprised of a number of subsystems, are uncorrelated then we require that the
total correlation of the $\left\{  AB\right\}  $ system is simply the sum of
the correlation within $A$ and $B$ separately.

Previously we have extended this measure to examine the properties of
multipartite correlations and entanglement [4]. Characterizing the
entanglement of multipartite systems is non-trivial [5], but the information
content of the total correlation is a fundamental measure that can be employed
to yield useful general properties of the correlation. In this note we explore
the use of this parameter to investigate the properties of multipartite
correlation where the end-goal is to establish an information-theoretic bound
on entanglement swapping.

\section{Partitioning}

The notion of partitioning of quantum systems is a key idea that we will make
use of extensively. In order to illustrate this we consider 4 qubits labelled
1,2,3 and 4. There are various, equivalent, ways we can approach this system.
We could, for example, consider qubits 2,3 and 4 to be a single `system'. We
could then consider the correlation between qubit 1 and qubits 2,3 and 4 taken
as a single system. Where several quantum systems are grouped together in this
fashion we use the notation $\left\{  234\right\}  $ to emphasize this
grouping. In this case we term the correlation between qubit 1 and the system
of qubits $\left\{  234\right\}  $ as an `external' correlation denoted by
$E_{1\left\{  234\right\}  }$.

The reason for this nomenclature is that the total information content of the
correlation between the 4 qubits is simply given by%
\begin{align}
I_{1234}  & =S_{1}+S_{2}+S_{3}+S_{4}-S_{1234}\nonumber\\
& =\left(  S_{2}+S_{3}+S_{4}-S_{234}\right)  +\left(  S_{1}+S_{234}%
-S_{1234}\right) \nonumber\\
& =I_{234}+E_{1\left\{  234\right\}  }%
\end{align}
where $I_{234}$ can be interpreted as the correlation `internal' to the
$\left\{  234\right\}  $ system of qubits so that the total correlation is
just the sum of the internal correlation plus the external correlation. This
is a general property of this entropic measure of correlation. For example,
the total correlation for these 4 qubits can also be written as%
\begin{equation}
I_{1234}=I_{12}+I_{34}+E_{\left\{  12\right\}  \left\{  34\right\}  }%
\end{equation}
where here we interpret the total system as comprising 2 sub-systems $\left\{
12\right\}  $ and $\left\{  34\right\}  $. The total correlation is again the
sum of the internal correlations $I_{12}$ and $I_{34}$, and the external
correlation $E_{\left\{  12\right\}  \left\{  34\right\}  }$.

It is important to note two things. Firstly, the partitioning is only notional
although we could conceive of constructing such partitions physically (for
example, we could physically separate qubits 1 and 2 from qubits 3 and 4).
Secondly, the quantities $I$ and $E$ have the same mathematical form being the
sum of sub-system entropies minus the total entropy. We distinguish them to
emphasize the partitioning into internal and external.

This notion of external and internal correlation is the same as that employed
in the Ithaca interpretation of quantum mechanics [6,7] where the notion that
correlation between systems is sufficient to describe their properties is
developed. Its utility here is that certain partitions have invariant
correlation under unitary transformation.

This partitioning extends, in some sense, to the interactions between the
sub-systems. For example, if we consider our 4 qubits to be interacting
unitarily with one another then we would have interaction terms in the
Hamiltonian of the form $\hat{H}_{jk}$. If we consider the partitioning into
qubit 1 and qubits $\left\{  234\right\}  $ then we would have some effective
interaction Hamiltonian $\hat{H}_{1\left\{  234\right\}  }^{eff}$ that
describes the evolution of the total system. As we have already mentioned,
this partitioning is useful in allowing us to construct invariant entropies
and correlations in a unitary interaction between quantum systems.

Entanglement swapping is a special case of quantum teleportation [8] and the
phenomenon has been experimentally demonstrated [9]. In the usual entanglement
swapping scheme we begin with 2 pairs of entangled particles. So we might
consider the qubits $\left\{  12\right\}  $ to be initially entangled and the
qubits $\left\{  34\right\}  $ to be initially entangled, with no entanglement
or correlation between the $\left\{  12\right\}  $ and $\left\{  34\right\}  $
partitions. We let qubits 2 and 3 interact unitarily and perform a measurement
(or equivalently we perform a Bell measurement on qubits 2 and 3). This
procedure results in qubits 1 and 4 becoming entangled with one another. The
central feature here is that qubits 2 and 3 interact. Accordingly in this note
we wish to study the evolution of the correlations when the sub-system
components are allowed to interact unitarily.

\section{Two Interacting Systems}

For two quantum systems, which we label as 1 and 2, the information content of
the correlation is given as above in (1). This quantity tells us the
difference in information between considering the systems 1 and 2 separately
and considering them as one entity $\left\{  12\right\}  $. If systems 1 and 2
are themselves comprised of sub-systems then equation (1) gives us the
`external' correlation between 1 and 2. This can, of course, be cast more
formally in terms of the Hilbert spaces. So system 1 might be described by
states in the space $H_{1}=H_{a}\otimes H_{b}\otimes H_{c}\ldots$ and system 2
by states in the space $H_{2}=H_{\alpha}\otimes H_{\beta}\otimes H_{\gamma
}\ldots$ where the subscripts refer to individual quantum systems such as qubits.

The correlation is bounded by [2]%
\begin{equation}
I_{12}\leq2\inf\left\{  S_{1},S_{2}\right\}
\end{equation}
and if $\left\{  12\right\}  $ is in a pure state then $S_{1}=S_{2}$. Let us
suppose that these systems are prepared in some initial state (which could be
mixed) and we let 1 and 2 interact unitarily then the correlation is
time-dependent, but the total entropy remains invariant and we have that%
\begin{equation}
I_{12}\left(  t\right)  =S_{1}\left(  t\right)  +S_{2}\left(  t\right)
-S_{12}\left(  0\right)
\end{equation}
Noting that the sum of the individual entropies must always be less than or
equal to the sum of the individual maximum entropies so that $S_{1}\left(
t\right)  +S_{2}\left(  t\right)  \leq S_{1}^{\max}+S_{2}^{\max}$ we have that%
\begin{equation}
I_{12}\left(  t\right)  \leq S_{1}^{\max}+S_{2}^{\max}-S_{12}\left(  0\right)
\end{equation}
If the 2 systems are initially uncorrelated and maximally mixed so that
$S_{12}\left(  0\right)  =S_{1}^{\max}+S_{2}^{\max}$ then it is easy to see
that the interaction cannot develop any correlation between the systems.
Conversely, if the two systems are initially in a pure state and maximally
correlated the interaction can only reduce that correlation.

These two special instances are well-known and obvious properties of
correlations for these initial states but they illustrate the general approach
we shall take here. We consider unitarily interacting sub-systems and examine
the entropy and correlation invariants of that interaction in order to yield
general properties for the evolution of the correlations.

\section{Three Interacting Systems}

As a precursor to the situation relevant to entanglement swapping we now
consider 3 quantum systems labelled 1,2 and 3. We shall assume, for
convenience, these systems are not comprised of internal sub-systems so that
we can set their `internal' correlation to zero. We shall further assume that
system 3 is initially uncorrelated with the system described by the $\left\{
12\right\}  $ partition. System 3 interacts unitarily with system 2 for a time
$t$. \ The total $\left\{  123\right\}  $ system thus evolves unitarily
according to $\hat{U}=\hat{I}_{1}\otimes\hat{U}_{23}$ where $\hat{I}_{1}$ is
the indentity operator for the subspace of system 1 and $\hat{U}_{23}$ is the
unitary interaction between 2 and 3. We now ask the following questions. How
does the interaction affect the total degree of correlation $I_{123}$? How
does the interaction between 2 and 3 affect the correlation $I_{12}$ between 1
and 2?

\subsection{Entropy and Correlation Invariants}

Since systems 2 and 3 interact unitarily certain entropies, and hence
correlations, are invariant. For example, considering the entropy of system 1
we note that no local unitary operation on $\left\{  23\right\}  $ will change
this and so $S_{1}$ is time-invariant. Considerations of this sort allow us to
write down the following invariant entropies%
\begin{align}
S_{1}\left(  t\right)   & =S_{1}\left(  0\right) \nonumber\\
S_{23}\left(  t\right)   & =S_{23}\left(  0\right) \nonumber\\
S_{123}\left(  t\right)   & =S_{123}\left(  0\right)
\end{align}
All other entropies being time-dependent. Considering the correlation between
system 1 and the partition $\left\{  23\right\}  $ which is given by the
external correlation $E_{1\left\{  23\right\}  }=S_{1}+S_{23}-S_{123}$ then is
is clear that this external correlation is also time-invariant so that
\begin{equation}
E_{1\left\{  23\right\}  }\left(  t\right)  =E_{1\left\{  23\right\}  }\left(
0\right)
\end{equation}
where this latter condition merely expresses the fact that for 2 quantum
systems $A$ and $B$ no local unitary operation on $B$ will affect the degree
of entanglement between $A$ and $B$. The entropies $S_{2}$ and $S_{3}$ are
clearly time-dependent since $\rho_{2}$ and $\rho_{3}$ undergo non-unitary evolutions.

\subsection{The Total Correlation}

Using the notion of partitioning the total correlation can be decomposed into
external and internal as follows%
\begin{equation}
I_{123}\left(  t\right)  =I_{23}\left(  t\right)  +E_{1\left\{  23\right\}
}\left(  t\right)
\end{equation}
where here we recall that we have assumed no internal correlation for the
individual sub-systems so that $I_{1}=0$. The external correlation is
invariant and therefore simply equal to the initial correlation between 1 and
2 (since we have assumed system 3 is initially uncorrelated). The total
correlation is therefore%
\begin{equation}
I_{123}\left(  t\right)  =I_{123}\left(  0\right)  +I_{23}\left(  t\right)
\end{equation}
which gives the appealing and intuitive result that the total correlation is
simply the sum of the initial correlation and the correlation that develops
between 2 and 3 as a result of their interaction. Since we have $I_{23}\left(
t\right)  \geq0$ we can also see that the interaction between 2 and 3 can only
increase the total correlation (or leave it unchanged).

\subsection{The Correlation Between 1 and 2}

The total correlation for the 3 systems can be written in two equivalent ways
by considering the partition into 1 and $\left\{  23\right\}  $ and the
partition into $\left\{  12\right\}  $ and 3 so that%

\begin{align}
I_{123}\left(  t\right)   & =I_{23}\left(  t\right)  +E_{1\left\{  23\right\}
}\left(  t\right) \nonumber\\
& =I_{12}\left(  t\right)  +E_{3\left\{  12\right\}  }\left(  t\right)
\end{align}
A simple rearrangement gives us that%
\begin{equation}
E_{1\left\{  23\right\}  }\left(  t\right)  -I_{12}\left(  t\right)
=E_{3\left\{  12\right\}  }\left(  t\right)  -I_{23}\left(  t\right)
\end{equation}
However, $E_{1\left\{  23\right\}  }\left(  t\right)  $ is an invariant so
that $E_{1\left\{  23\right\}  }\left(  t\right)  =E_{1\left\{  23\right\}
}\left(  0\right)  =I_{12}\left(  0\right)  $ which gives%
\begin{equation}
I_{12}\left(  0\right)  -I_{12}\left(  t\right)  =E_{3\left\{  12\right\}
}\left(  t\right)  -I_{23}\left(  t\right)
\end{equation}
Strong subadditivity [9] gives us the condition that $E_{3\left\{  12\right\}
}\left(  t\right)  \geq I_{23}\left(  t\right)  $ and so we obtain the result
that $I_{12}\left(  0\right)  -I_{12}\left(  t\right)  \geq0$. This
establishes the following theorem

\begin{quote}
If we have 3 quantum systems 1,2 and 3 such that 3 is initially uncorrelated
with either 1 or 2 and we let 2 interact unitarily with 3, then the
interaction always reduces the correlation between 1 and 2, or leaves it unchanged.
\end{quote}

This is consistent with the monogamy property of quantum mechanics in which
maximal pairwise entanglement can not be established for more than 1 pair of a
3 component system [10]. Although in the above we have only considered systems
with no degree of internal correlation for convenience, this result is easily
extended to the case where 1,2, and 3 are each comprised of a number of sub-systems.

Of course the content of this theorem is intuitive and obvious; if we have
systems 1 and 2 with some initial degree of correlation then we would not
expect some local process on 2 to \textit{increase} the degree of correlation.
However, as the next example shows we must sometimes be careful in relying on
our intuition where correlation is concerned.

\subsection{Non-Transitivity of Correlation}

If we consider 3 systems $A,B$ and $C$ then if $A$ is correlated with $B$ and
$B$ is correlated with $C$ then it would seem intuitive to suppose that $A$
has to be correlated to $C$. This, however, is not always true as the
following counter-example shows.

Let us suppose that $A$ is a single qubit, qubit 1, system $B$ is comprised of
2 qubits, qubits 2 and 3 and $C$ is a single qubit, qubit 4. If we prepare
these qubits in the state%

\begin{equation}
\left\vert \psi\right\rangle =\frac{1}{\sqrt{2}}\left(  \left\vert
0\right\rangle _{1}\left\vert 0\right\rangle _{2}+\left\vert 1\right\rangle
_{1}\left\vert 1\right\rangle _{2}\right)  \otimes\frac{1}{\sqrt{2}}\left(
\left\vert 0\right\rangle _{3}\left\vert 0\right\rangle _{4}+\left\vert
1\right\rangle _{3}\left\vert 1\right\rangle _{4}\right)
\end{equation}
then it is easy to see that $I_{1B}\neq0,I_{4B}\neq0$ but $I_{14}=0$. Qubit 1
is correlated to qubit 2, and qubit 3 is correlated to qubit 4, but system $B$
is comprised of qubits 2 and 3 which are not correlated with one another. At
this `system' level, therefore, it is not possible to demonstrate a
transitivity property for correlation because internally the chain of
correlation can be broken within a given system.

The state given by (14) is, of course, that considered in typical
entanglement-swapping schemes where the qubit pairs are initially maximally
entangled. This state possesses the maximum possible total correlation
$I_{1234}$ for 4 qubits, even though the overall $\left\{  1234\right\}  $
system is not maximally entangled. In order to have maximal entanglement we
have to have a state that \textit{simultaneously} optimizes the pairwise
correlations [4]. The state (14) clearly does not simultaneously optimize the
pairwise correlation between the qubits since qubits 2 and 3 are uncorrelated
whereas the qubit pairs $\left\{  12\right\}  $ and $\left\{  34\right\}  $
are maximally correlated.

\subsection{Entanglement Exchange in an Atom-Field Interaction}

In the usual entanglement-swapping scheme there are 4 qubits such that
$\left\{  12\right\}  $ are entangled and $\left\{  34\right\}  $ are
entangled with no entanglement between these partitions so that $E_{\left\{
12\right\}  \left\{  34\right\}  }\left(  0\right)  =0$. The initial
entanglement is `swapped' by ineracting qubits 2 and 3 followed by a
subsequent measurement on these qubits. The result is that the initial
entanglement can be transferred to qubits 1 and 4, which have never previously
interacted. Entanglement swapping can be viewed as an instance of quantum
teleportation [11].

The key feature here is that the meaurement can be viewed as projecting the
$\left\{  14\right\}  $ qubits into an entangled state. The measurement
process is, of course, non-unitary. It is, however, possible to exchange
entanglement to 2 qubits that have never directly interacted using only
unitary processes. To illustrate this we consider an idealized example
consisting of 2 two-level atoms interacting with a single field mode in a
lossless cavity [12]. The appropriate interaction Hamiltonian is given by the
Jaynes-Cummings Hamiltonian in the rotating-wave approximation%
\begin{equation}
\hat{H}_{int}=\hat{a}\hat{\sigma}_{+}+\hat{a}^{\dag}\hat{\sigma}_{-}%
\end{equation}
where $\hat{a},\hat{a}^{\dag}$ are the field annihilation and creation
operators, respectively, and $\hat{\sigma}_{-},\hat{\sigma}_{+}$ are the
atomic lowering and raising operators, respectively.

We consider the first atom to be prepared in its excited state and the field
to be in its vacuum state. The first atom is sent through the cavity with a
cavity transit time such that there is a probability of 1/2 of the atom-field
interaction resulting in the atom leaving the cavity in its ground state. The
state of the total system after this interaction is therefore given by a state
of the form%
\begin{equation}
\left\vert \psi\right\rangle =\frac{1}{\sqrt{2}}\left(  \left\vert
0\right\rangle _{A_{1}}\left\vert 1\right\rangle _{F}+\left\vert
1\right\rangle _{A_{1}}\left\vert 0\right\rangle _{F}\right)  \otimes
\left\vert \varphi\right\rangle _{A_{2}}%
\end{equation}
where the subscripts $A$ and $F$ refer to the atoms and field, respectively.
We consider this first interaction to be a state preparation phase that
generates an initial correlation between atom 1 and the field. If we now
consider the second atom to be prepared in its ground state and sent through
the cavity with a transit time such that there would be a unit probability of
the atom absorbing the photon if the field were in the state $\left\vert
1\right\rangle _{F}$ then the total state after this interaction is given by%
\begin{equation}
\left\vert \psi\right\rangle =\frac{1}{\sqrt{2}}\left(  \left\vert
0\right\rangle _{A_{1}}\left\vert 1\right\rangle _{A_{2}}+\left\vert
1\right\rangle _{A_{1}}\left\vert 0\right\rangle _{A_{2}}\right)
\otimes\left\vert 0\right\rangle _{F}%
\end{equation}
The field state has been completely `decoupled' by the second interaction and
the atom-field entanglement after the first interaction has been transferred
to an entanglement between the 2 atoms. In our general notation and
terminology above, atom 1 would be system 1, the field system 2, and atom 2
would be system 3. Thus the unitary interaction of 2 and 3 has `decoupled'
system 2 and the initial entanglement between 1 and 2 transferred to an
entanglement between 1 and 3.

The entropy of atom 1 during this second interaction remains unchanged; the
degree of mixing of the state of atom 1 (after the first preparation
interaction) is invariant when the field interacts with atom 2. The initial
correlation that exists between atom 1 and the field is reduced and the
correlation between atom 1 and atom 2 increases. We have chosen interaction
times to generate maximally entangled states here, but it is straightforward
to generalize this to abitrary cavity transit times for the atoms. The
correlations that develop between the atoms and field are consistent with the
general properties (10) and (13) above.

\section{Four Quantum Systems}

We now consider the situation most pertinent to entanglement swapping where we
have 4 quantum systems, labelled 1,2,3 and 4. Here, however, we are going to
consider a more general scenario in which these 4 quantum systems can each be
comprised of a number of sub-systems. The usual entanglement swapping scheme
is just a special case of this more general situtation.

As before we shall let systems 2 and 3 interact unitarily and we shall also
assume that there is no initial external correlation between the $\left\{
12\right\}  $ and $\left\{  34\right\}  $ partitions\footnote{Since systems
1,2 3 and 4 are themselves possibly comprised of sub-systems then these can
also be viewed as partitions of the total system. The partition $\left\{
12\right\}  $ is really then a partition of partitions although we shall
simply refer to it as the $\left\{  12\right\}  $ partition.}. This can be
represented schematically as%
\[%
\begin{array}
[c]{ccc}%
1\Diamond &  & \Diamond4\\
\uparrow &  & \uparrow\\
\downarrow &  & \downarrow\\
2\Diamond & \longleftarrow\hat{U}_{23}\longrightarrow & \Diamond3
\end{array}
\]
This is a quite general model for interacting systems. If we wish to consider
the interaction of 2 systems prepared in mixed states, for example, then 1 and
4 might be taken to be the supplementary quantum systems required for the
purification of the $\left\{  12\right\}  $ and $\left\{  34\right\}  $
partitions. Our goal here, as in the previous section, is to examine the
general properties of the correlations that develop as a result of the interaction.

It is clear from the previous section that the unitary interaction between 2
and 3 will reduce any initial external correlation between 1 and 2 (or at best
leave it unchanged) so that $E_{12}\left(  t\right)  \leq E_{12}\left(
0\right)  $. It is also clear, from symmetry, that the initial external
correlation between 3 and 4 will also be reduced by the interaction (or at
best unchanged) so that $E_{34}\left(  t\right)  \leq E_{34}\left(  0\right)
$.

\subsection{Entropy and Correlation Invariants}

Since the interaction between 2 and 3 is assumed to be unitary then there are
various entropies, and hence correlations, that remain invariant under the
interaction. There are 7 invariant entropies and these are; $S_{1234}%
,S_{1},S_{4},S_{23},S_{14},S_{123},$ and $S_{234}$. All other entropies are
time-dependent. There are 30 possible correlations we can consider; the 4
internal correlations $I_{k}$, the correlations internal to a given partition,
and the various external correlations between the partitions. There are 6
invariant correlations from the 30 possibilities. For our purposes we shall
consider only the following 4 invariants%
\begin{align}
I_{1}\left(  t\right)   & =I_{1}\left(  0\right) \nonumber\\
I_{4}\left(  t\right)   & =I_{4}\left(  0\right) \nonumber\\
E_{1\left\{  234\right\}  }\left(  t\right)   & =E_{1\left\{  234\right\}
}\left(  0\right) \nonumber\\
E_{\left\{  23\right\}  4}\left(  t\right)   & =E_{\left\{  23\right\}
4}\left(  0\right)
\end{align}

\subsection{The Total Correlation}

The total correlation can be partitioned as%

\begin{align}
I_{1234}\left(  t\right)   & =I_{1}\left(  t\right)  +I_{\left\{  234\right\}
}\left(  t\right)  +E_{1\left\{  234\right\}  }\left(  t\right) \nonumber\\
& =const+I_{\left\{  234\right\}  }\left(  t\right)
\end{align}
where we have used the invariants (18). The internal correlation $I_{\left\{
234\right\}  }\left(  t\right)  $ can also be partitioned as $I_{\left\{
234\right\}  }\left(  t\right)  =I_{4}\left(  t\right)  +E_{\left\{
23\right\}  4}\left(  t\right)  +I_{23}\left(  t\right)  $. The quantity
$I_{23}\left(  t\right)  $ is just the correlation that develops between 2 and
3 as a result of their interaction. Since $I_{4}$ and $E_{\left\{  23\right\}
4}$ are invariant, the total correlation can be written as
\begin{equation}
I_{1234}\left(  t\right)  =const+I_{23}\left(  t\right)  =I_{1234}\left(
0\right)  +I_{23}\left(  t\right)
\end{equation}
where for the latter identity we have assumed that 2 and 3 are initially
uncorrelated. We therefore arrive at the following general theorem

\begin{quote}
If $A$ and $B$ are 2 initially uncorrelated quantum systems each comprised of
a number of sub-systems and a partition of $A$ interacts unitarily with a
partition of $B$ then the total correlation that develops is greater than or
equal to the total initial correlation and is simply the sum of the initial
correlation and the correlation that develops between the partitions.
\end{quote}

This, again, is an intuitive and appealing general result. It is interesting
that the interaction reduces certain correlations between systems (the initial
external correlation between 1 and 2 reduces, for example) but in such a way
that the total correlation increases if the initial correlation is not maximal.

In an optimal entanglement swapping scheme qubits $\left\{  12\right\}  $ are
maximally entangled, as are qubits $\left\{  34\right\}  $. In this case it is
clear that $I_{23}\left(  t\right)  =0$ so that no correlation develops
between 2 and 3 as a result of their (unitary) interaction. It is only when
the initial internal correlations for the partitions $\left\{  12\right\}  $
and $\left\{  34\right\}  $ are not maximal will there be any correlation
developed between qubits 2 and 3. This gives us the seemingly paradoxical
property that it is only when no correlation develops between 2 and 3 (which
implies that qubits 2 and 3 are in maximally mixed states) can we transfer
maximal entanglement to qubits 1 and 4.

\section{Entanglement Swapping}

We now consider the general problem of entanglement swapping for qubits. If
the qubit pairs are not initially maximally entangled, then what is the
maximum correlation, or entanglement, that can be swapped? A general\ and
elegant approach to this in terms of concurrence has been developed [13], but
here we show how a simple information-theoretic bound can be established.

In the previous sections we have considered a unitary evolution of the state;
in entanglement swapping a measurement is necessary to transfer the
entanglement to the 2 qubits that have not directly interacted. Consider 4
qubits, (labelled 1,2,3,4 from left to right, as necesasary), prepared in the state%

\begin{equation}
\left\vert \psi\right\rangle =\left(  a\left\vert 00\right\rangle +b\left\vert
11\right\rangle \right)  \otimes\left(  c\left\vert 00\right\rangle
+d\left\vert 11\right\rangle \right)
\end{equation}
which we can represent diagramatically as%
\[
\fbox{$\Diamond~\leftrightarrows~\Diamond$}~~\otimes~~\fbox{$\Diamond
~\leftrightarrows~\Diamond$}
\]
We denote the index of correlation before the entanglement swapping process
with $I$ and afterwards by $I^{M}$ where the superscript reminds us that we
are considering the situation before and after the Bell measurement. Writing
the Bell basis in the usual fashion as%
\begin{align}
\left\vert \Psi_{\pm}\right\rangle  & =\frac{1}{\sqrt{2}}\left(  \left\vert
00\right\rangle \pm\left\vert 11\right\rangle \right) \nonumber\\
& \nonumber\\
\left\vert \Phi_{\pm}\right\rangle  & =\frac{1}{\sqrt{2}}\left(  \left\vert
01\right\rangle \pm\left\vert 10\right\rangle \right)
\end{align}
our initial state of the 4 qubits can be written as%
\begin{align}
\left\vert \psi\right\rangle _{1234}  & =\frac{1}{\sqrt{2}}\left(
ac\left\vert 00\right\rangle _{14}+bd\left\vert 11\right\rangle _{14}\right)
\otimes\left\vert \Psi_{+}\right\rangle _{23}+\frac{1}{\sqrt{2}}\left(
ac\left\vert 00\right\rangle _{14}-bd\left\vert 11\right\rangle _{14}\right)
\otimes\left\vert \Psi_{-}\right\rangle _{23}\nonumber\\
& \nonumber\\
& +\frac{1}{\sqrt{2}}\left(  ad\left\vert 01\right\rangle _{14}+bc\left\vert
10\right\rangle _{14}\right)  \otimes\left\vert \Phi_{+}\right\rangle
_{23}+\frac{1}{\sqrt{2}}\left(  ad\left\vert 01\right\rangle _{14}%
-bc\left\vert 10\right\rangle _{14}\right)  \otimes\left\vert \Phi
_{-}\right\rangle _{23}\nonumber\\
& \nonumber\\
&
\end{align}
Defining the normalized states%
\begin{align}
\left\vert \psi_{\pm}\right\rangle  & =n_{\psi}^{-1/2}\left(  ac\left\vert
00\right\rangle \pm bd\left\vert 11\right\rangle \right) \nonumber\\
\left\vert \varphi_{\pm}\right\rangle  & =n_{\varphi}^{-1/2}\left(
ad\left\vert 01\right\rangle \pm bc\left\vert 10\right\rangle \right)
\end{align}
with $n_{\psi}=a^{2}c^{2}+b^{2}d^{2}$ and $n_{\varphi}=a^{2}d^{2}+b^{2}c^{2}$
(where the modulus has been dropped for convenience) we can write the initial
state as%
\begin{equation}
\left\vert \psi\right\rangle _{1234}=\frac{1}{\sqrt{2}}n_{\psi}^{1/2}\left(
\left\vert \psi_{+}\right\rangle \left\vert \Psi_{+}\right\rangle +\left\vert
\psi_{-}\right\rangle \left\vert \Psi_{-}\right\rangle \right)  +\frac
{1}{\sqrt{2}}n_{\varphi}^{1/2}\left(  \left\vert \varphi_{+}\right\rangle
\left\vert \Phi_{+}\right\rangle +\left\vert \varphi_{-}\right\rangle
\left\vert \Phi_{-}\right\rangle \right)
\end{equation}
where the states appearing to the left side of the tensor products describe
the $\left\{  1,4\right\}  $ qubits and those to the right the $\left\{
2,3\right\}  $ qubits. So that a Bell measurement on the $\left\{
2,3\right\}  $ qubits projects the $\left\{  1,4\right\}  $ qubits into the
states%
\begin{align*}
& \left\vert \psi_{\pm}\right\rangle \text{ \ \ with probability \ \ }\frac
{1}{2}n_{\psi}\\
& \\
& \left\vert \varphi_{\pm}\right\rangle \text{ \ \ with probability \ \ }%
\frac{1}{2}n_{\varphi}%
\end{align*}

\subsection{Upper Bound on Correlation}

Let us assume that the Bell measurement on the $\left\{  2,3\right\}  $ qubits
has been performed with the result $\left\vert \Psi_{+}\right\rangle $
obtained. This result is communicated to the holders of the $\left\{
1,4\right\}  $ qubits. The $\left\{  1,4\right\}  $ qubits can then be
assigned the pure state%
\begin{equation}
\left\vert \psi_{+}\right\rangle =n_{\psi}^{-1/2}\left(  ac\left\vert
00\right\rangle +bd\left\vert 11\right\rangle \right)
\end{equation}
The density operator for qubit 1 is therefore%
\begin{align}
\hat{\rho}_{1}^{M}  & =\frac{a^{2}c^{2}}{n_{\psi}}\left\vert 0\right\rangle
\left\langle 0\right\vert +\frac{b^{2}d^{2}}{n_{\psi}}\left\vert
1\right\rangle \left\langle 1\right\vert \nonumber\\
& \nonumber\\
& =\left(  \frac{1}{2}+\varepsilon_{1}^{M}\right)  \left\vert 0\right\rangle
\left\langle 0\right\vert +\left(  \frac{1}{2}-\varepsilon_{1}^{M}\right)
\left\vert 1\right\rangle \left\langle 1\right\vert
\end{align}
where $\left\vert \varepsilon_{1}^{M}\right\vert $ is the bias. A qubit state
of higher entropy has a lower bias and vice versa.

Let us assume without loss of generality that the pre-measurement indices of
correlation satisfy $I_{12}\geq I_{34}$ which implies that $c^{2}\geq a^{2}$
and that both $a^{2},b^{2}>d^{2}$. We further assume, again without any
essential loss of generality, that $a^{2}>b^{2}$. The pre-measurement biases
for qubits 1 and 3 are therefore%
\begin{align}
\varepsilon_{1}  & =a^{2}-1/2\nonumber\\
\varepsilon_{3}  & =c^{2}-1/2
\end{align}
with the post-measurement bias for qubit 1 being given by%
\begin{align}
\varepsilon_{1}^{M}  & =\frac{a^{2}c^{2}}{a^{2}c^{2}+b^{2}d^{2}}-\frac{1}%
{2}\nonumber\\
& \nonumber\\
& =c^{2}\left(  \frac{1}{c^{2}+d^{2}\left(  b^{2}/a^{2}\right)  }\right)
-\frac{1}{2}%
\end{align}
but
\begin{equation}
\left(  \frac{1}{c^{2}+d^{2}\left(  b^{2}/a^{2}\right)  }\right)  >1
\end{equation}
Hence $\varepsilon_{1}^{M}>\varepsilon_{3}$. This implies that the entropy of
qubit 1 after measurement is lower than the entropy of qubit 3 before
measurement and so we have $I_{14}^{M}\leq I_{34}$. Similar arguments apply to
all possible output states after the measurement and so we have the result
that%
\begin{equation}
\fbox{$I_14^M\leq\inf\left\{  I_12,I_34\right\}  $}%
\end{equation}
The degree of entanglement that can be transferred is therefore limited by the
lower of the initial existing degrees of entanglement.

\subsection{Example}

With the choices $a^{2}=3/4$ and $c^{2}=7/8$ and assuming the result of the
Bell measurement is $\left\vert \Psi_{+}\right\rangle $ then the density
operator for qubit 1 post-measurement is%
\begin{equation}
\hat{\rho}_{1}^{M}=\frac{21}{22}\left\vert 0\right\rangle \left\langle
0\right\vert +\frac{1}{22}\left\vert 1\right\rangle \left\langle 1\right\vert
\end{equation}
which is a good deal less mixed than the pre-measurement density operators for
the qubits. Note that any iteration of the entanglement swapping process in
which we begin with less than perfect entanglement will rapidly drive the
$\left\{  1,4\right\}  $ qubits into an uncorrelated state.

\section{Conclusions}

Entanglement remains one of the most intriguing features of quantum mechanics.
Indeed, many have argued that it is \textit{the} central feature of quantum
mechanics that distinguishes it from a classical perspective. Characterizing
the entanglement of multipartite systems is a difficult, and still largely
unresolved, problem. In this note we have emphasized a measure of correlation
based on the information content of the correlation. For bi-partite systems in
a pure state this is just proportional to the entropy of entanglement. The
generalization of this to multipartite systems that we have used here does not
provide a similarly straightforward measure of entanglement. It does, however,
give a useful measure of the overall degree of correlation within any given
partition of a quantum system and between those partitions. It is an
observable-independent characterization that provides the unique measure
satisfying the additivity property that if two quantum systems, each comprised
of sub-systems, are uncorrelated then the total correlation is simply the sum
of the correlations \textit{within} those two multi-component systems.

In this note we have used this measure, together with the notion of
partitioning, to derive some general properties of the correlation of
interacting quantum systems. Partitioning is only notional unless we take
steps to physically create the partitions, but it allows us to identify the
various entropy and correlation invariants of the interaction. Once these
invariants have been identified it only requires very elementary techniques to
establish these general properties. In particular, it is easy to demonstrate
using this approach the intuitive result that if we have 2 initially
uncorrelated quantum systems \textit{A} and \textit{B}, each comprised of a
number of sub-systems, and a partition of \textit{A} interacts unitarily with
a partition of \textit{B}, then the total correlation is simply the sum of the
initial correlation \textit{within} \textit{A} and \textit{within} \textit{B}
and the time-dependent correlation due to the interaction. It would certainly
be surprising if it were otherwise, but the index of correlation applied to
multipartite systems allows us to quantify this precisely in
information-theoretic terms.

These general results apply only to unitary interactions between the
partitions. In entanglement swapping a non-unitary process (i.e. measurement)
is employed to transfer the entanglement to 2 systems that have never
previously directly interacted. Viewing entanglement as a resource, two remote
parties can make use of this entanglement provided they are given the
supplementary information about the result of the measurement. The index of
correlation allows us to place an upper bound to the amount of entanglement
that can be transferred if the initial systems are not prepared in perfectly
entangled states. \ The maximum amount of entanglement that can be transferred
in entanglement swapping is the lower of the 2 initial entanglements.

\section{Acknowledgements}

I would like to thank S.M. Barnett and S. Croke for helpful and illuminating discussions.

\section{References}

\begin{enumerate}
\item D. Janzing, \textquotedblleft Entropy of Entanglement\textquotedblright%
,. in D. Greenberger, K. Hentschel and F. Weinert, (eds), \textit{Compendium
of Quantum Physics}. Springer, 205 (2009)

\item S.M. Barnett and S.J.D. Phoenix, \textquotedblleft Information Theory,
Squeezing and Quantum Correlations\textquotedblright, \textit{Phys. Rev. A},
\textbf{44} 535 (1991)

\item S.J.D .Phoenix and F.S. Khan, \textquotedblleft Partitions of Correlated
Quantum Systems\textquotedblright, in preparation (2016)

\item S.J.D. Phoenix, \textquotedblleft Quantum Information as a Measure of
Correlation\textquotedblright, \textit{Quant. Inf. Comp}, \textbf{14(10)} 3723 (2015)

\item R. Horodecki, P. Horodecki, M. Horodecki, and K. Horodecki,
\textquotedblleft Quantum Entanglement\textquotedblright, \textit{Rev. Mod.
Phys}. \textbf{81}, 865 (2009)

\item N.D. Mermin, \textquotedblleft The Ithaca Interpretation of Quantum
Mechanics\textquotedblright, https://arxiv.org/abs/quant-ph/9609013 (1996)

\item N.D. Mermin, \textquotedblleft What is quantum mechanics trying to tell
us?\textquotedblright, \textit{Am. J. Phys.} \textbf{66}, 753-767 (1998)

\item C. H. Bennett, G. Brassard, C. Cr\'{e}peau, R. Jozsa, A. Peres, and W.K.
Wootters, \textquotedblleft Teleporting an Unknown Quantum State via Dual
Classical and Einstein-Podolsky-Rosen Channels\textquotedblright,
\textit{Phys. Rev. Lett}. \textbf{70}, 1895, (1993)

\item M. Halder, A. Beveratos, N. Gisin, V. Scarani, C. Simon, and H. Zbinden,
\textquotedblleft Entangling independent photons by time
measurement\textquotedblright\ \textit{Nature Phys}, \textbf{3}, 692 (2007)

\item S.M. Barnett, \textquotedblleft Quantum Information\textquotedblright,
\textit{Oxford University Press}: Oxford, (2009)

\item W. K. Wootters, \textquotedblleft Entanglement of Formation of an
Arbitrary State of Two Qubits\textquotedblright, \textit{Phys. Rev. Lett}.
\textbf{80}, 2245 (1998)

\item S.J.D. Phoenix \& S.M. Barnett, \textquotedblleft Nonlocal Interatomic
Correlations in the Micromaser\textquotedblright, \textit{J. Mod. Opt}.,
\textbf{40} 979 (1993)

\item B. T. Kirby, S. Santra, V. S. Malinovsky, and M. Brodsky,
\textquotedblleft Entanglement swapping of two arbitrarily degraded entangled
states\textquotedblright, \textit{Phys. Rev. A} \textbf{94}, 012336 (2016)
\end{enumerate}

\end{document}